\documentclass{iaus} 
\usepackage{graphicx} 
 
\title[Secular Evolution in Disk Galaxies] 
{Internal Secular Evolution in Disk Galaxies:\\The Growth of Pseudobulges} 
 
\author[short author list]   
{John Kormendy} 
 
\affiliation{Department of Astronomy, University of Texas at Austin, USA;
      \break Max-Planck-Institut f\"ur Extraterrestrische Physik, Garching, Germany;
      \break Universit\"ats-Sternwarte, Munich, Germany
      \break email: kormendy@astro.as.utexas.edu
}
 
\pubyear{2007} 
\volume{245}  
\pagerange{000--000} 
\date{?? and in revised form ??} 
\jname{Formation and Evolution of Galaxy Bulges} 
\editors{M. Bureau, et al. eds.} 
\begin{document} 
 
\maketitle


\def\ts{\thinspace}
\def\etal{{\it et~al.~}} \def\etnuk{{\it et~nuk.~}}
\def\gapprox{$_>\atop{^\sim}$} \def\lapprox{$_<\atop{^\sim}$}
\def\kms{km~s$^{-1}$}
\newdimen\sa  \def\sd{\sa=.1em \ifmmode $\rlap{.}$''$\kern -\sa$
                               \else \rlap{.}$''$\kern -\sa\fi}
\newdimen\sb  \def\md{\sb=.02em\ifmmode $\rlap{.}$'$\kern -\sb$
                               \else \rlap{.}$'$\kern -\sb\fi}
\def\msun {M$_{\odot}$~} \def\msund{M$_{\odot}$}
\def\mbh{$M_{\bullet}$~} \def\mbhd{$M_{\bullet}$}
\def\m31{M{\ts}31} \def\mm32{M{\ts}32} 
\def\vs{\vskip 0pt}\def\vss{\vskip 0pt}
\def\0{\phantom{0}}
\def\ba{\kern -1pt}
\def\x{\ts\ba$\times$\null}
\def\dot{\0\ts\raise 0.2em\hbox{{\dots}}}
\def\cl{\centerline}
\def\nhi{\noindent \hangindent=15pt}

\shortauthor{Kormendy}
\shorttitle{Internal Secular Evolution in Galaxies}

\begin{abstract} 
      Observational and theoretical evidence that internal, slow (``secular'') 
evolution reshapes galaxy disks is reviewed in Kormendy \& Kennicutt (2004).  
This update has three aims.  First, I emphasize that this evolution is very 
general~-- it is as fundamental to the evolution of galaxy disks as
(e.{\ts}g.)~core collapse is to globular clusters, as the production of 
hot Jupiters is to the evolution of protoplanetary disks, and as evolution
to red giants containing proto-white-dwarfs is to stellar evolution.
One consequence for disk galaxies is the buildup of dense central 
components that get mistaken for classical (i.{\ts}e., merger-built) 
bulges but that were grown out of disk stars and gas.  We call these 
pseudobulges.  Second, I review new results on pseudobulge star formation 
and structure and on the distinction between boxy and disky pseudobulges.
Finally, I highlight how these results make a galaxy formation problem more
acute.  How can hierarchical
clustering produce so many pure disk galaxies with no evidence for 
merger-built bulges?

\end{abstract}

\pretolerance=10000   \tolerance=10000

\section{The fundamental way that disks evolve is by spreading.}

      Galactic evolution is in transition from the early Universe dominated by
hierarchical clustering to a future dominated by internal secular evolution. 
There are many ways that stars and gas can interact with collective phenomena 
such as bars, oval disks, spiral structure, and triaxial dark halos.  
Different processes deserve{\ts}--{\ts}and are getting{\ts}--{\ts}detailed study. 
In this paper, I want to emphasize common features that make the ensemble of 
internal evolution processes fundamental.  This point is important enough that
it should appear in these Proceedings, so I quote 
a discussion from Kormendy \& Fisher (2005):

      ``A general principle
of the evolution of self-gravitating systems is that it is energetically favorable 
to spread -- to shrink the inner parts by expanding the outer parts.  How to 
see this depends on whether the system is dominated by rotation or by random
motions.

\subsection{If Dynamical Support Is By Random Motions}

\def\ts{\thinspace}

      Then the argument (Lynden-Bell \& Wood 1968; Binney \& Tremaine 1987) is 
based on the fundamental point that the specific heat of a self-gravitating
system is negative.  Consider an equilibrium system of $N$ particles of mass
$m$, radius~$r$, and three-dimensional velocity dispersion~$v$.~The virial
theorem says that 2{\ts}KE + PE = 0, where the kinetic energy KE = $Nmv^2/2$
and the potential energy PE = $-G(Nm)^2/r$ define $v$~and~$r$. The total energy
of a bound system, $E \equiv$ KE $+$ PE = $-$KE, is negative.  But temperature
$T$ corresponds to internal velocity as $mv^2/2 = 3kT/2$.  So the specific heat 
$C \equiv dE/dT \propto d(-Nmv^2/2)/d(v^2)$ is also negative.  In the above, $G$ is
the gravitational constant and $k$ is Boltzmann's~constant.

      The system is supported by heat, so evolution is by heat transport.
If the center of the system gets hotter than the periphery, then heat tends to flow 
outward.  The inner parts shrink and get still hotter.  This promotes further
heat flow. The outer parts receive heat; they expand and cool.  Whether the
system evolves on an interesting timescale depends on whether there is an
effective heat-transport mechanism.  For example, many globular clusters evolve
quickly by two-body relaxation and undergo core collapse.  Giant 
elliptical galaxies -- which otherwise would evolve similarly -- cannot do so
because their relaxation times are much longer than the age of the Universe.

\subsection{If Dynamical Support Is By Rotation}

      Tremaine (1989) provides a transparent summary of an argument due to
Lynden-Bell \& Kalnajs (1972) and to Lynden-Bell \& Pringle (1974).  A disk is
supported by rotation, so evolution is by angular momentum transport.  The
`goal' is to minimize the total energy at fixed total angular momentum.  
A rotationally supported ring at radius $r$ in a fixed potential $\Phi(r)$ has
specific energy $E(r)$ and specific angular momentum $L(r)$ given by
$$E(r) = {r\over2}{d\Phi \over dr} + \Phi~~{\rm and}~~
L(r) = \biggl(r^3 \ts{d\Phi \over dr}\biggr)^{1/2}~.$$
Then $dE/dL = \Omega(r)$, where $\Omega = (r^{-1} d\Phi/dr)^{1/2}$ is the
angular speed of rotation.  Disks spread when a unit mass at radius $r_2$
moves outward by gaining angular momentum $dL$ from a unit mass at 
radius $r_1 < r_2$. This is energetically favorable: the change in energy, 
$$
dE = dE_1 + dE_2 = \biggl[- \biggl({dE \over dL}\biggr)_1 
                          + \biggl({dE \over dL}\biggr)_2\ts\biggr]dL 
                 = \ts[-\Omega(r_1) + \Omega(r_2)]\ts dL\ts,
$$
is negative because $\Omega(r)$ usually decreases outward.  `Thus disk
spreading leads to a lower energy~state.  In general, disk spreading,
outward angular momentum flow, and energy dissipation accompany one another
in astrophysical disks' (Tremaine 1989).  

\subsection{Self-Gravitating Systems Evolve By Spreading}

      The consequences are very general.  All of the following are caused by
the above physics.

      Globular and open clusters are supported by random motions, so they 
spread in three dimensions by outward energy transport.  The mechanism is
two-body relaxation, and the consequences are core collapse and the evaporation
of the outer parts.

      Stars are spherical systems supported by pressure.  They spread in three
dimensions by outward energy transport.  The mechanisms are radiation or
convection mediated by opacity.  Punctuated by phases of stability when nuclear
reactions replace the energy that is lost, stellar evolution consists of a
series of core contractions and envelope expansions.  One result is red
(super)giants containing cores that will become white dwarfs, neutron stars,
or stellar mass black holes.

      Protostars are spherical systems coupled to circumstellar disks by
magnetic fields that wind up because of differential rotation. This drives
jets that look one-dimensional but that really are three-dimensional; they
carry away angular momentum  and allow the inner circumstellar disk to shrink 
and accrete onto the star (Shu et al.~1994, 1995). 

      Protoplanetary disks are supported by rotation; they spread in two
dimensions by outward angular momentum transport.  Dynamical friction
produces, for example, hot Jupiters and colder Neptunes.

      Galactic disks are supported by rotation.  They want to spread in
two dimensions by outward angular momentum transport.  Efficient driving
mechanisms are provided by bars and globally oval disks.
Like all of the above, the evolution is secular -- it is slow compared to
the collapse time of the disk.  Secular evolution is the subject of this paper.''

\begin{figure*}[b]
\includegraphics[height=4.85cm]{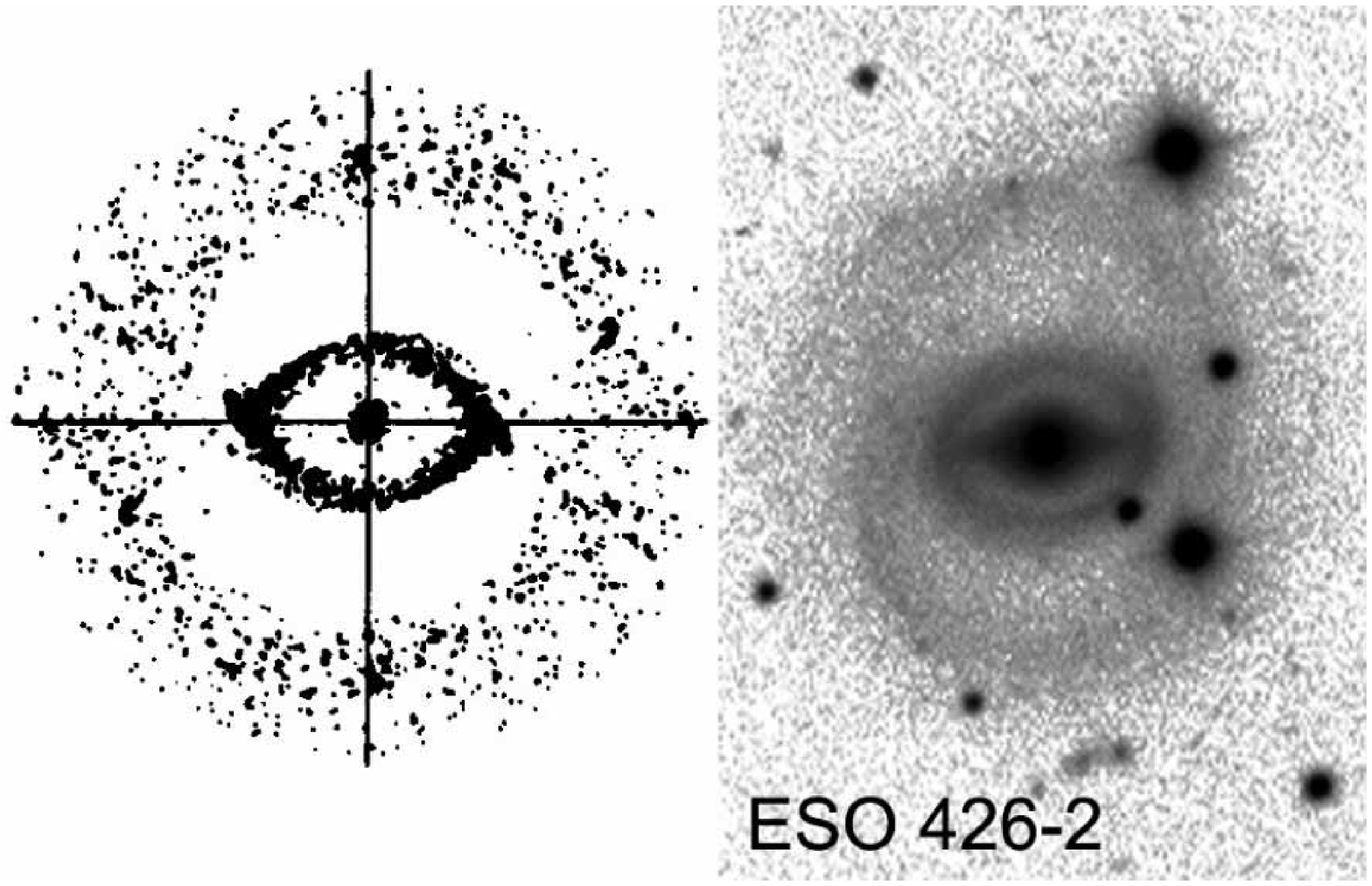}%
\hspace*{0.1cm}%
\includegraphics[height=4.85cm]{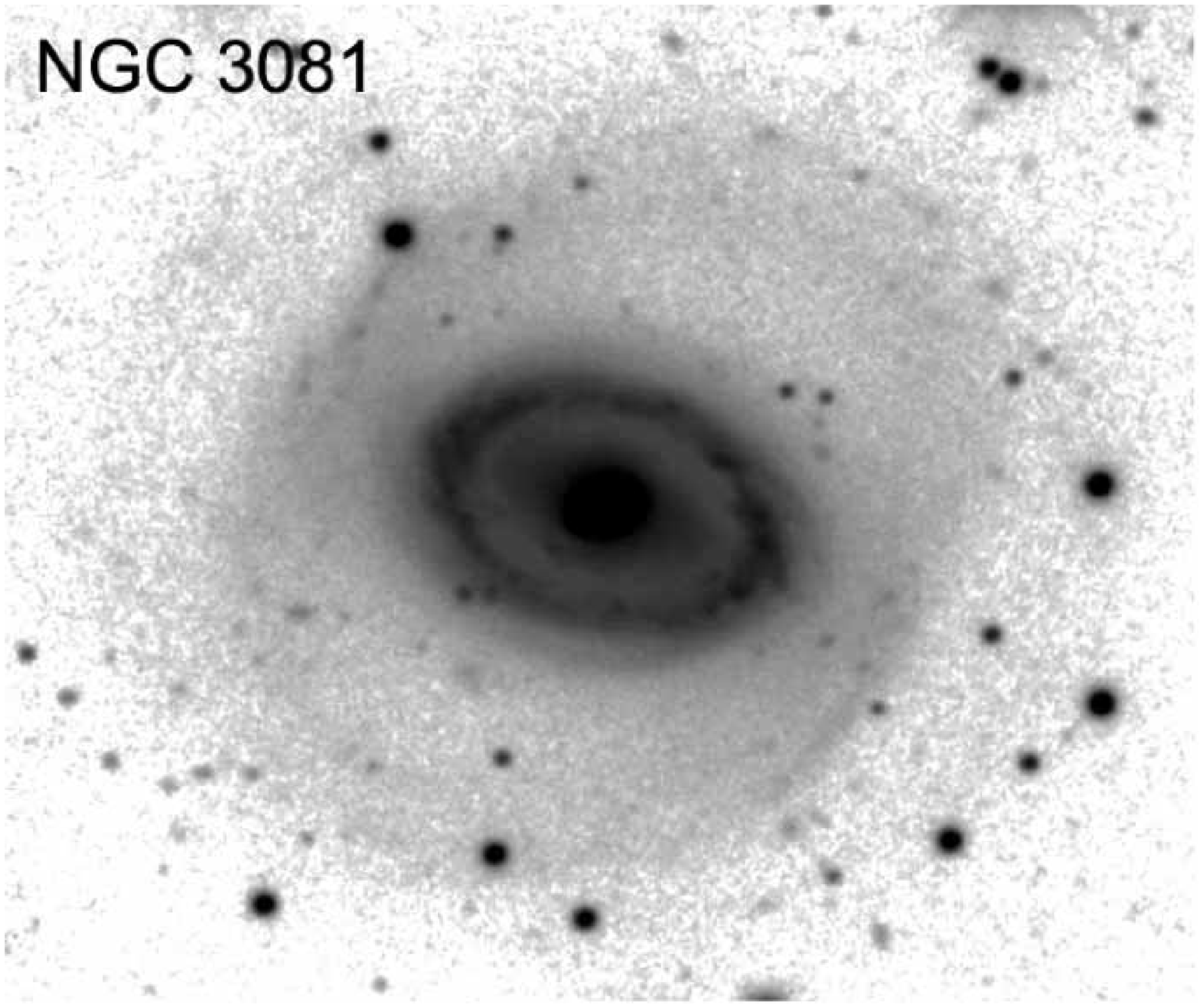}
\caption{Secular evolution products: (left) Gas particles at the end of a 
sticky-particle simulation of evolution in a rotating bar potential 
(horizontal but not shown; Simkin, Su, \& Schwarz 1980).  After 7 bar rotations, 
gas has collected into an outer ring, an inner ring around the end of the bar, 
and a dense central concentration.  Such features are seen in SB galaxies;
e.{\ts}g., ESO 426-2 (Buta \& Crocker 1991) and NGC 3081 (Buta, Corwin, \& Odewahn 
2007).  Detailed hydrodynamic simulations confirm and enrich this picture
(see especially Athanassoula 1992).  This figure is adapted from 
Kormendy \& Kennicutt (2004).}
\label{fig:dummy}
\end{figure*}

\section{Secular Evolution and the Growth of Pseudobulges}

      That interactions with collective phenomena such as bars and spiral 
structure result in secular evolution was emphasized as long ago as Kormendy 
(1979a,{\ts}b; 1982a,{\ts}b).  One of the earliest 
examples of a specific process was the suggestion by Duus \& Freeman (1975), 
now confirmed by many papers (e.{\ts}g., Simkin, Su, \& Schwarz 1980),
that gas in a barred galaxy settles into a ring surrounding the end 
of the bar (Figure 1).  The formation of bars is likely to be 
partly secular (Sellwood 2000).  Another early result 
was the demonstration (Combes \& Sanders 1981) that $n$-body bars 
heat up in the axial direction; when seen edge-on, they resemble -- and, 
we believe, explain -- box-shaped bulges (\S\ts4).  By 1982, the secular
evolution of barred galaxies was a thriving industry (see Kormendy 1982a for 
a review).  Sellwood \& Wilkinson (1993), Kormendy (1993), and Buta \& Combes 
(1996) provide interim reviews.  In the past decade, progress 
has accelerated rapidly.  Recent reviews include 
Kormendy \& Cornell (2004);
Kormendy \& Fisher (2005);
Athanassoula (2007a, b), and most thoroughly, 
Kormendy \& Kennicutt (2004, hereafter KK04).

       Figure 1 illustrates the results of gas evolution in a barred galaxy.
Disk gas is rearranged into an ``outer ring'' at $\sim 2.2$ bar radii, 
an ``inner ring'' that encircles the end of the bar, and a dense central
concentration of gas.  As the gas density increases, star formation 
is likely, and indeed, the features produced in \underbar{gas} closely resemble 
the \underbar{stellar} outer rings, inner rings, and pseudobulges seen in the
figure.  Much observational evidence supports these interpretations, 
as discussed in the above reviews.

     Pseudobulges are the consequence discussed in this paper.
In the simulations, gas falls to the center and builds high densities, 
often in rings.  Since star formation rate density $\Sigma_{\rm SFR}$ increases 
faster than linearly with gas density $\Sigma_{\rm gas}$, 
$\Sigma_{\rm SFR} \propto \Sigma_{\rm gas}^{1.4}$
(Kennicutt 1998a, b), high star formation rates are expected.
They are observed.  KK04 review
observations of nuclear starbursts, often in spectacular
rings and often associated with bars and 
oval disks. The star formation rates in starbursting nuclear rings
imply, with modest replenishment of the observed nuclear gas, 
that they would build the stellar densities that we observe in 
pseudobulges in 1 -- 3 billion years.  
That is, the formation picture suggested by the simulations is clearly 
consistent, via observed gas densities, star formation rates, and plausible 
timescales, with the disky pseudobulges discussed in the next section.

\section{Update on Pseudobulge Properties and Star Formation}

     The properties that allow us to classify pseudobulges are listed
in KK04.  I have space 
here for only a brief update of pseudobulge properties and star formation rates.

     Fisher (2006) used {\it Spitzer Space Telescope\/} images to derive, for
50 galaxies, 3.6 $\mu$m - 8 $\mu$m color profiles that are a measure of star
formation rates (Wu~et~al. 2005). He found that morphologically identified
classical bulges have lower star formation rates than their disks.  Pseudobulges
have star formation rates similar to those of their associated disks.  If most 
pseudobulges show ongoing star formation, then it must be secular (KK04).

     Drory \& Fisher (2007a, b) show that bulge type correlates with 
the division of galaxies into a red sequence and a blue cloud 
in Sloan Survey color-magnitude diagrams (Strateva et al.~2001).  Red sequence galaxies 
contain mainly classical bulges.  Blue cloud galaxies contain pseudobulges.
The division into red and blue is not due to different mixtures
of red bulges and blue disks; rather, classical bulge galaxies \underbar{globally}
formed most of their stars early, and pseudobulge galaxies (excepting S0s) 
continue to form stars throughout. 

      Peletier (2007) reviews SAURON observations that show, in superb, two-dimensional
detail, evidence for pseudobulges with disky dynamics and young stellar populations.

      Fisher \& Drory (2007) measure surface brightness distributions
in 84, S0{\ts}--{\ts}Sc galaxies in which they morphologically classify the bulges
as classical or pseudo.  They decompose major-axis profiles into S\'ersic (1968) 
$\log{I} \propto r^{1/n}$ profiles $+$ exponential disks.  They find:

\begin{enumerate}

\item Almost all pseudobulges have $n \leq 2$ and almost all classical bulges have 
      $n \geq 2$.  This extends results of (e.{\ts}g.)
      Courteau, de Jong, \& Broeils (1996); 
      Carollo et al.~(2002); 
      MacArthur, Courteau, \& Holtzman (2003), and
      KK04.  
      We do not understand how to predict $n$ in either type of bulge, but the above 
      correlation is good enough so we can use $n$ as an (imperfect, to be sure) 
      classification criterion.

\item Bulge-to-total luminosity ratios $B/T$ are smaller for 
      pseudobulges than for classical bulges over the whole Hubble
      sequence and also at a given Hubble type.  Almost no pseudobulges
      have $B/T > 1/3$, as expected if they are built out of disk material.

\item Pseudobulges are flatter than classical bulges; in fact, the flattest classical 
      bulge in their sample is less flattened than the average pseudobulge.  But the 
      roundest classical and pseudo bulges have similar intrinsic flattenings.  
      Not all pseudobulges are flat. 

\end{enumerate}

$\phantom{000}$ \vskip -40pt $\phantom{000}$

\section{Where are the bulges in galaxies with box-shaped, edge-on bars?}

      Figure 2 (left) shows a normal, early-type barred galaxy.  It is easy to 
distinguish the bar from the bulge (which happens to be a pseudobulge: KK04).  
Ignoring, for present purposes, the distinction between lens and outer ring, 
there are three main components, the disk, the bar, and the pseudobulge.  The 
next two panels show NGC 4565, a familiar galaxy with a boxy pseudobulge.  The 
middle image is an infrared (3.6 $\mu$m) version of the view that we normally see 
in print.  Two components are visible, the disk and the boxy bulge.  As long as we 
thought that a boxy shape was a minor structural detail of bulges, our mental 
picture of this galaxy was entirely canonical: it is an Sb galaxy with a big disk 
and a small bulge.  But now we know that boxy bulges are edge-on bars.  So the 
central image shows two components, a disk and a bar.  Where is the bulge in this 
Sb galaxy?  The same question applies to all edge-on galaxies with boxy bulges. 

      The right-hand image and plot provide the answer (Kormendy \& Barentine 2007).
NGC 4565 contains a high-surface-brightness but tiny pseudobulge that is clearly 
distinct from the boxy bar. 
The key to seeing this component is to observe far enough into the infrared to reduce
the dust absorption that is a problem in the optical.  A 3.6\ts$\mu$m {\it Spitzer Space 
Telescope\/} archive image and an HST NICMOS image are unaffected by dust near the center.  
The latter provides high enough
spatial resolution so that we can measure the S\'ersic index of the central component.
Is it a classical bulge or is it a pseudobulge?~It's axial ratio is consistent with
a classical bulge -- it is not particularly flat.  But it's S\'ersic index is
$n = 1.33 \pm 0.12$. This means that it is a pseudobulge (\S\ts3).  Kormendy \& 
Barentine (2007) find similar results for NGC 5746, another edge-on galaxy with a boxy bulge.

     This is remarkable:  Except for the faint thick disk and 
halo, {\it the thickest component in NGC 4565 is the edge-on bar, which is part of the 
disk.  The next thickest component is the thin disk.  The component with the smallest 
scale height is the pseudobulge.}  It has the shape of a classical bulge, but its 
scale height is only $1\sd2 \sim 90$ pc.  In contrast, the scale height of the 
boxy bar $+$ thin disk is $10^{\prime\prime} \sim 790$ pc.
Four implications deserve emphasis:

\begin{enumerate}

\item Seeing a pseudobulge distinct from the boxy bar increases confidence 
in our picture of secular evolution.  It is easier to believe that we understand boxy 
bars in edge-on galaxies if we also find (pseudo)bulges like those associated with bars 
in face-on galaxies.

\item $B/T$ ratios in edge-on galaxies with ``boxy bulges'' are smaller than we thought.

\item Published $B/T$ values in edge-on and face-on barred galaxies are inconsistent. 
In edge-on galaxies, we count box-shaped structures as bulge light.  When we see such 
a galaxy face-on, we identify this light as a bar and measure a smaller $B/T$ ratio. 

\item A problem with cold dark matter galaxy formation gets more
acute.  Simien \& de Vaucouleurs (1986) find that $B/T = 0.4$ in NGC 4565.
But $B$ refers to the boxy bar.  Figure 2 shows that the pseudobulge is much less 
luminous than the boxy structure.  And it has the properties of pseudobulges that 
were grown out of disks.  Yet the disk rotates at $255 \pm 10$ km s$^{-1}$ interior 
to the outer warp (Rupen 1991).  In a hierarchically clustering Universe, how can 
a galaxy grow so massive with no evidence for a major merger?

\end{enumerate}

\begin{figure*}[t]
\includegraphics[height=5.00cm]{kormendy_fig2a.eps}
\includegraphics{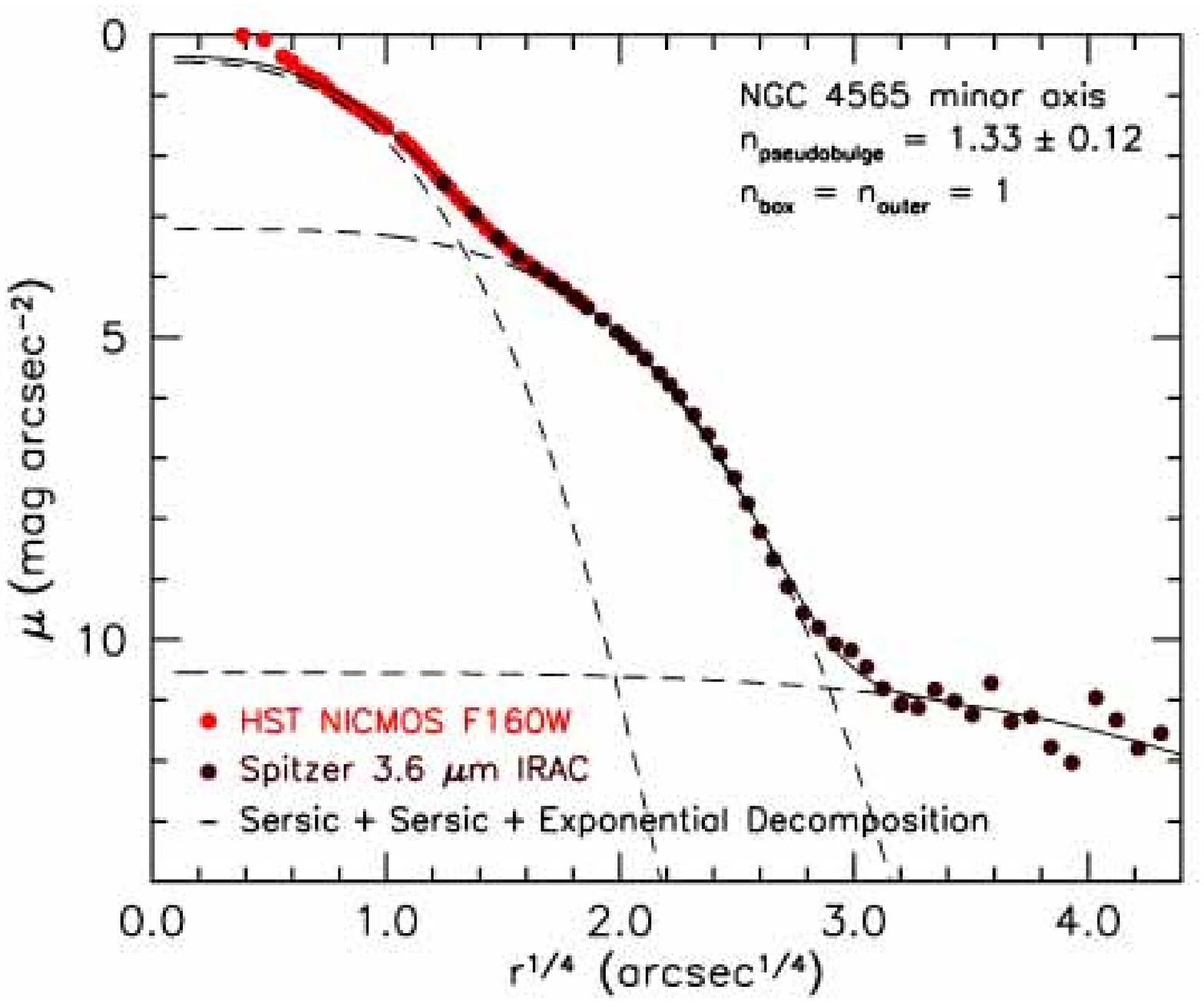} 
\includegraphics{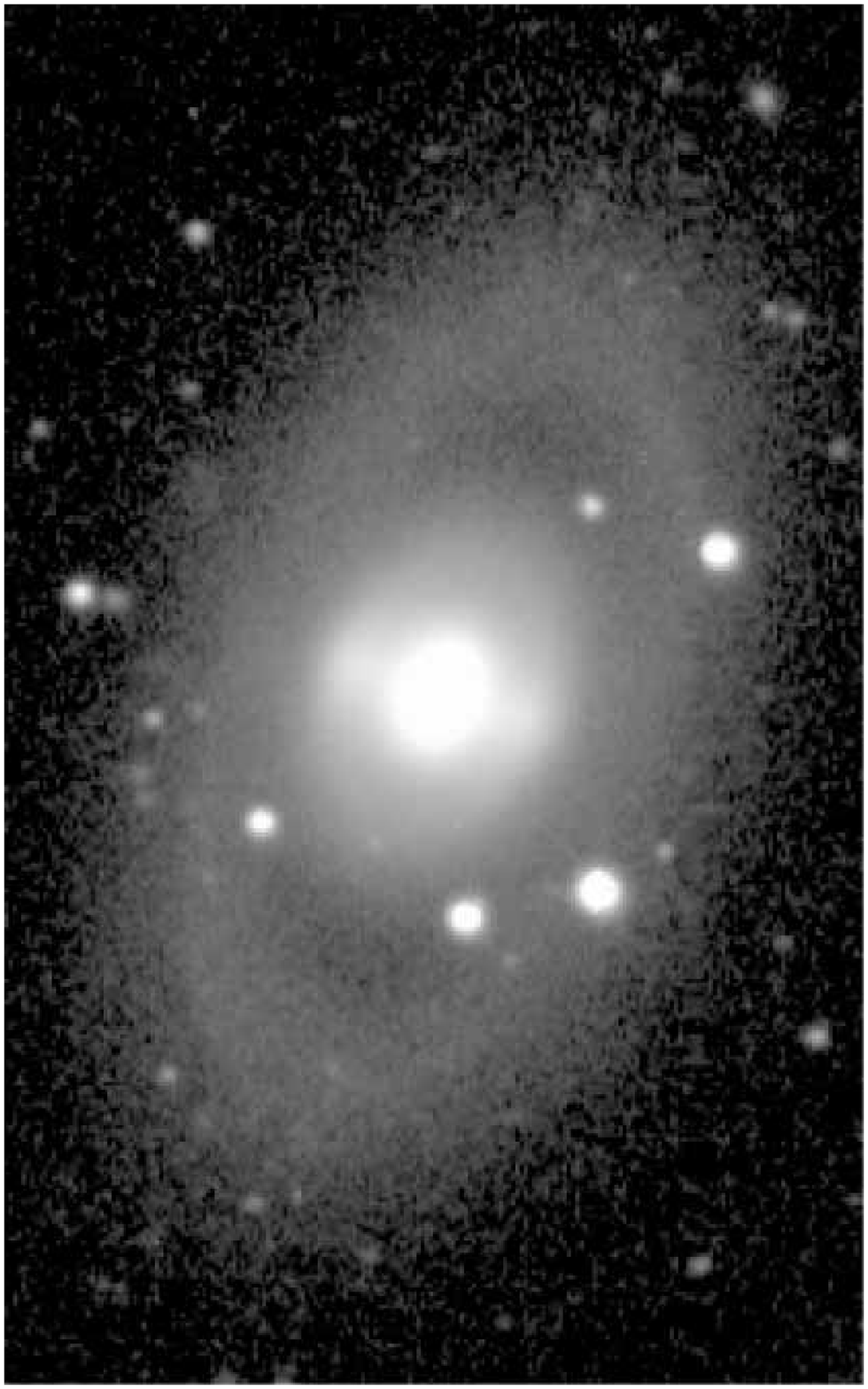} 
\includegraphics{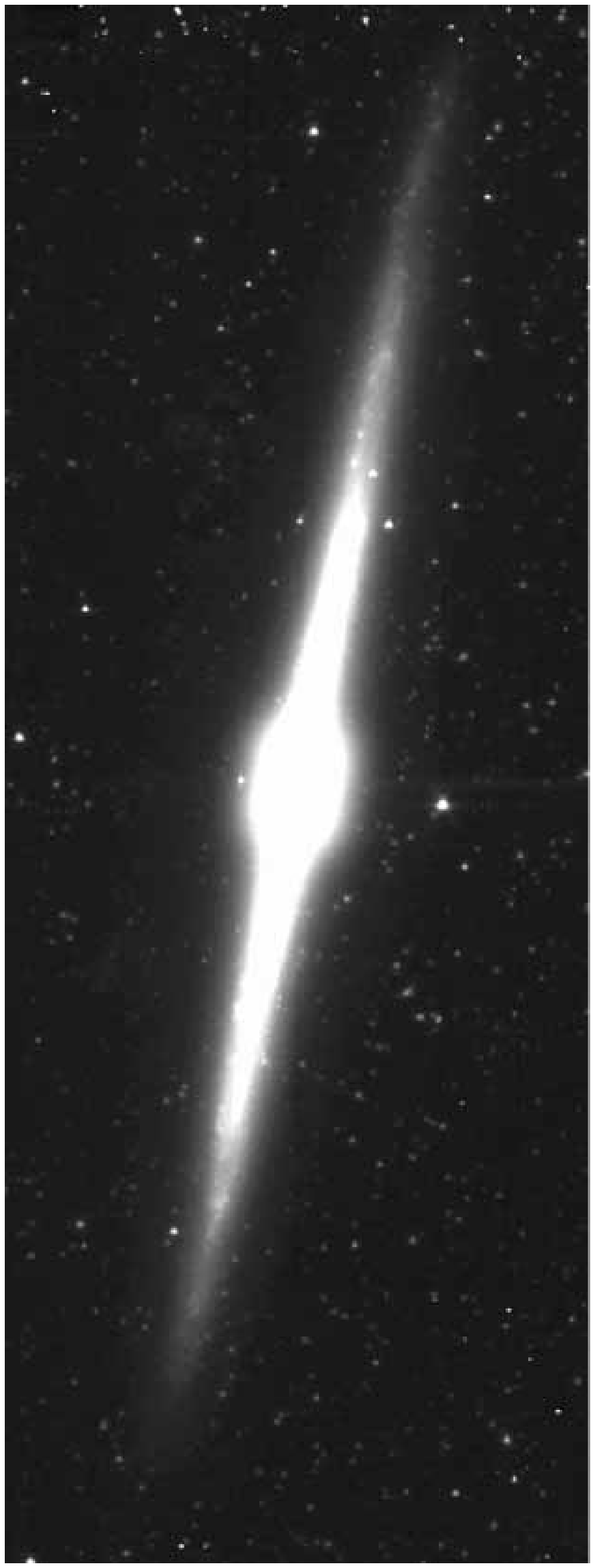} 
\includegraphics{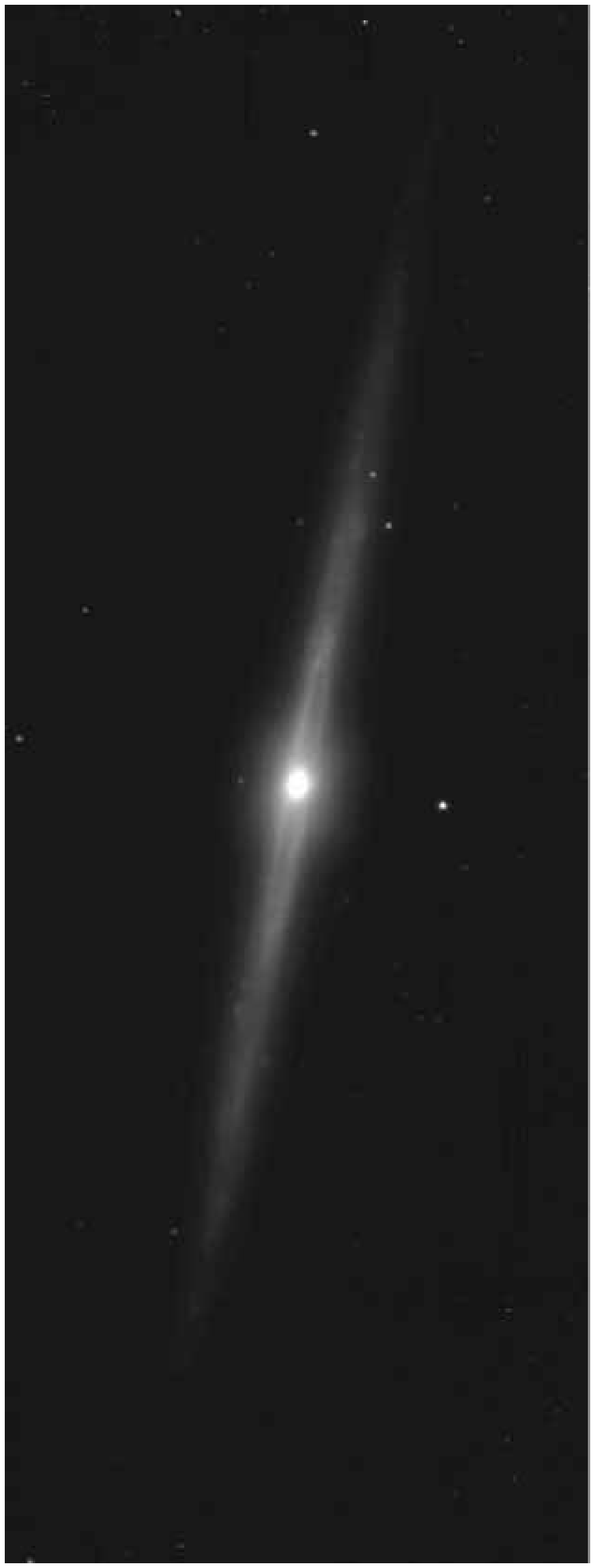} 
\caption{The left panel shows a visible-light image of the SB0 galaxy
         NGC 3945.  It contains a pseudobulge, a bar that fills a lens component along the
         minor axis, and an outer ring.  The central panels show a {\it Spitzer Space 
         Telescope\/} 3.6 $\mu$m IRAC image of NGC 4565 at different brightnesses and
         contrasts to emphasize the box-shaped bulge (left) and the 
         central pseudobulge (right).  The rightmost panel shows the minor-axis 
         brightness profile derived from the {\it Spitzer\/} image and from a 
         {\it Hubble Space Telescope\/} (HST) NICMOS F160W image.  Also shown is a 
         three-component decomposition into a central S\'ersic function for the pseudobulge, 
         a S\'ersic function for the boxy bulge, and an outer exponential.  
         From Kormendy \& Barentine (2007).}
\label{fig:dummy}
\end{figure*}

\begin{acknowledgments} 
      I thank John Barentine, Niv Drory, and David Fisher for permission to show results 
before publication.  David Fisher kindly provided the image of NGC 3945.  This paper 
used archive data from the {\it Hubble Space Telescope} and from the {\it Spitzer Space 
Telescope\/}.  The Space Telescope Science Institute is operated by AURA, Inc., and 
{\it Spitzer\/} is operated by the Jet Propulsion Laboratory (Caltech), both under contract 
with NASA.  This work was supported by the US National Science Foundation under grant
AST-0607490.
\end{acknowledgments}

\end{document}